\documentclass[sigconf]{acmart}

\AtBeginDocument{%
  \providecommand\BibTeX{{%
    \normalfont B\kern-0.5em{\scshape i\kern-0.25em b}\kern-0.8em\TeX}}}

\setcopyright{acmcopyright}
\copyrightyear{2018}
\acmYear{2018}
\acmDOI{XXXXXXX.XXXXXXX}

\acmConference[Conference acronym 'XX]{Make sure to enter the correct
  conference title from your rights confirmation emai}{June 03--05,
  2018}{Woodstock, NY}

\acmPrice{15.00}
\acmISBN{978-1-4503-XXXX-X/18/06}
\usepackage{multirow}
\begin{document}

\title{Attribute Simulation for Item Embedding Enhancement in Multi-interest Recommendation}


\author{Yaokun Liu}
\affiliation{%
  \institution{Tianjin University}
  \city{Tianjin}
  \country{China}
}
\email{yaokunl@tju.edu.cn}

\author{Xiaowang Zhang}
\affiliation{%
  \institution{Tianjin University}
  \city{Tianjin}
  \country{China}
}
\email{xiaowangzhang@tju.edu.cn}

\author{Minghui Zou}
\affiliation{%
  \institution{Tianjin University}
  \city{Tianjin}
  \country{China}
}
\email{minghuizou@tju.edu.cn}

\author{Zhiyong Feng}
\affiliation{%
  \institution{Tianjin University}
  \city{Tianjin}
  \country{China}
}
\email{zyfeng@tju.edu.cn}

\begin{abstract}

Although multi-interest recommenders have achieved significant progress in the matching stage, our research reveals that existing models tend to exhibit an under-clustered item embedding space, which leads to a low discernibility between items and hampers item retrieval. This highlights the necessity for item embedding enhancement. However, item attributes, which serve as effective and straightforward side information for enhancement, are either unavailable or incomplete in many public datasets due to the labor-intensive nature of manual annotation tasks. This dilemma raises two meaningful questions: 1. Can we bypass manual annotation and directly simulate complete attribute information from the interaction data? And 2. If feasible, how to simulate attributes with high accuracy and low complexity in the matching stage? 
 
In this paper, we first establish an inspiring theoretical feasibility that the item-attribute correlation matrix can be approximated through elementary transformations on the item co-occurrence matrix. Then based on formula derivation, we propose a simple yet effective module, \textbf{SimEmb} (Item \underline{Emb}edding Enhancement via \underline{Sim}ulated Attribute), in the multi-interest recommendation of the matching stage to implement our findings. By simulating attributes with the co-occurrence matrix, SimEmb discards the item ID-based embedding and employs the attribute-weighted summation for item embedding enhancement. Comprehensive experiments on four benchmark datasets demonstrate that our approach notably enhances the clustering of item embedding and significantly outperforms SOTA models with an average improvement of 25.59\% on Recall@20.
\end{abstract}

\vspace{-2cm}

\ccsdesc[500]{Information systems~Recommender systems}

\vspace{-2cm}

\keywords{Recommender Systems, Multi-interest Learning, Attribute Simulation, Item Embedding Enhancement}


\maketitle

\begin{figure}[t]
  \centering
  \includegraphics[width=\linewidth]{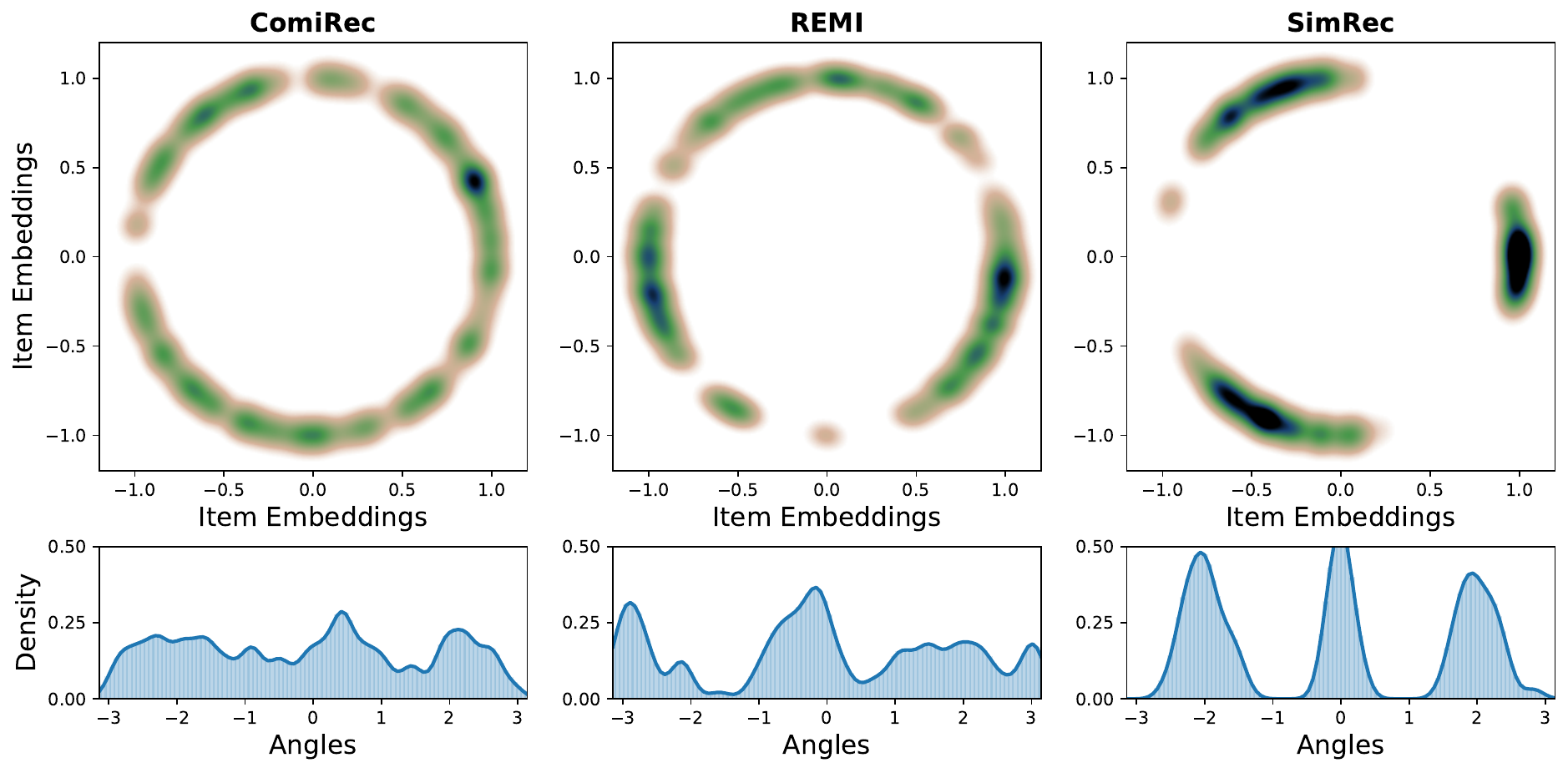}
  \caption{Item embedding distributions and density estimation curves of 600 items randomly sampled from RetailRocket. Item embeddings are mapped on a 2-dimensional unit circle with t-SNE \cite{t-sne} and KDE \cite{kde}, where darker colors indicate denser point clustering in this area. The density curve similarly displays point densities across angles. More details can be found in Section \ref{visualization}.}
  \label{cluster_intro}
  \vspace{-0.5cm}
\end{figure}

\section{INTRODUCTION}

In order to deliver personalized filtering for users amidst the vast pool of items on online platforms, industrial recommendation systems \cite{sdm, netflix} commonly follow a two-stage process: matching and ranking. The matching stage aims to extract a small subset of items from the extensive corpus with lightweight models, ensuring low computational costs. Subsequently, the ranking stage utilizes more sophisticated models to rerank the retrieved items. In this paper, we focus on improving the effectiveness of the matching stage, which serves as a crucial foundation for the recommendation system.

Recently, a rise of multi-interest learning-based methods has been witnessed in the sequential recommendation (SR) of the matching stage \cite{re4, udm}. The multi-interest recommendation extracts multiple interest vectors for each user based on their interaction sequences, effectively alleviating the information loss of using a single user embedding and gaining notable improvement in matching performance. MIND \cite{mind} pioneers using the capsule network with dynamic routing \cite{capsule} to capture multiple interests. Comirec \cite{comirec} introduces multi-head self-attention mechanisms and explores the balance between diversity and accuracy. Subsequent work \cite{pimirec, umi, remi} further improves from different aspects such as periodicity, user profile, and negative sampling.

\begin{figure}[t]
  \centering
  \setlength{\abovecaptionskip}{0.cm}
  \includegraphics[width=\linewidth]{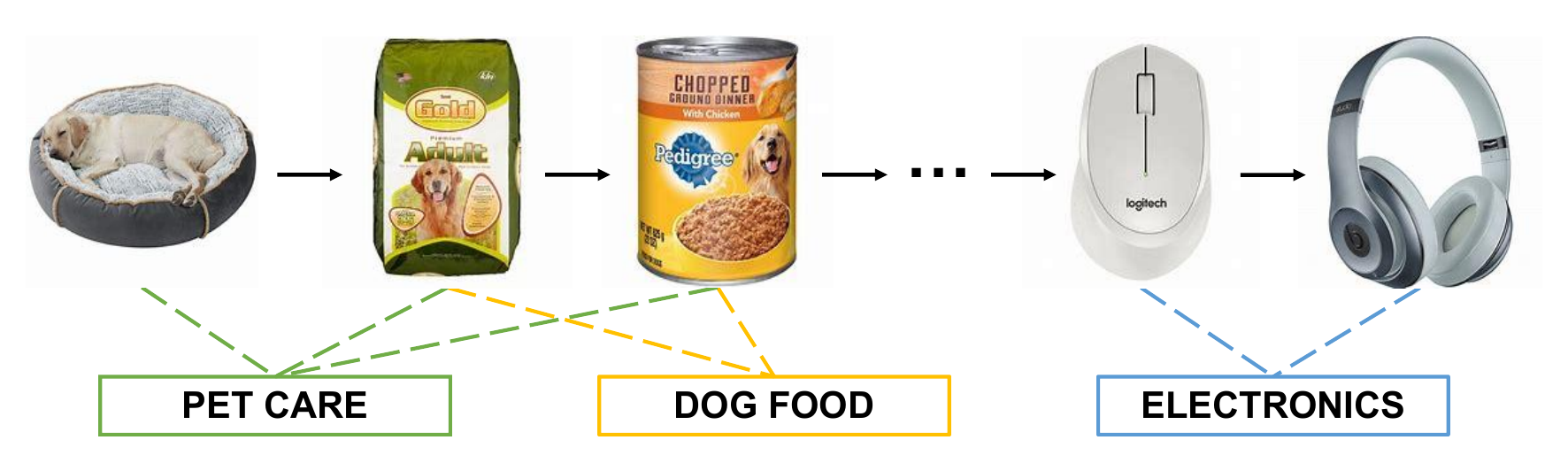}  
  \caption{An example of a user sequence and shared attributes between items.}
  \label{user_seq}
  \vspace{-0.5cm}
\end{figure}

Despite their effectiveness, our investigation reveals a clustering deficiency problem in the item embedding space of existing multi-interest models. As shown in Figure \ref{cluster_intro}, it is evident that the item embeddings of classic ComiRec and SOTA REMI models are uniformly distributed across the entire circle (in stark contrast to our proposed SimRec). This high similarity between different item embeddings hampers the item-matching efficiency, suggesting the necessity for item embedding enhancement.

To enhance item embeddings, an effective and straightforward strategy is to leverage item attribute side information, which provides rich semantic information. However, most multi-interest methods assume that only item IDs are available and do not take the item attribute into consideration. This situation is reasonable. As annotating item attributes is a time-consuming and laborious task, item attributes are either unavailable in many public datasets or sparse and incomplete, with numerous missing attribute values. It naturally leads to a meaningful question: \textit{Can we bypass the need for manual annotation of item attributes and, instead, simulate the attribute-item correlation from interaction data to achieve approximate effects?}

To answer this question, we find that the item co-occurrence information implicitly contains item-attribute correlations. As shown in Figure \ref{user_seq}, the co-occurrence of items within a user sequence strongly suggests specific similarities between them, implying the presence of shared attributes. To figure out this implicit containment relationship, we further conducted a theoretical analysis in Section \ref{theory}. The result demonstrates that the real item-attribute correlation matrix can be approximated by applying elementary transformations to the item co-occurrence matrix. 

However, although we establish a theoretical feasibility of simulating attributes which may inspire various implementation approaches, it is not trivial to apply it in the multi-interest learning of the matching stage, due to the difficulty of determining the correct transformation steps and the potential high complexity brought by the high dimensionality of the co-occurrence matrix. Consequently, a follow-up question emerges: \textit{Is there an implementation method to achieve both high simulation accuracy and low complexity?}

In this paper, we provide a positive answer to the question. Through formula derivation in Section \ref{derivation}, we discover that elementary transformations on the co-occurrence matrix and the attribute-weighted summation can merge and simplify into an intriguing equation, that is, the co-occurrence matrix multiplied by the attribute embedding matrix produces enhanced item embeddings. Based on this finding, we propose a simple yet effective module named \textbf{SimEmb} (Item \underline{Emb}edding Enhancement via \underline{Sim}ulated Attribute). Specifically, we first construct a normalized item co-occurrence matrix from user sequences. Then, we discard the existing item ID-based embedding layer and perform attribute-weighted summation based on the co-occurrence matrix to obtain the enhanced item embeddings. Finally, to further reduce the training complexity, we propose a process optimization of SimEmb.

The contributions of this paper can be summarized as follows:
\begin{itemize}
    \item We provide a theoretical feasibility that the item co-occurrence matrix can approximate the real item-attribute correlation matrix through elementary transformations.        
    \item We propose a simple yet effective item embedding module that enhances the item embedding through attribution simulation with the item co-occurrence matrix. 
    It can be an ideal alternative to the labor-intensive attribute annotating task.
    \item We conduct comprehensive experiments on four benchmark datasets, and the results demonstrate SimEmb significantly improves the effectiveness of various multi-interest models, with an average increase of 25.59\% on Recall@20.
\end{itemize}

\begin{figure*}[ht]
  \centering
  \setlength{\abovecaptionskip}{0.cm}
  \includegraphics[width=\linewidth]{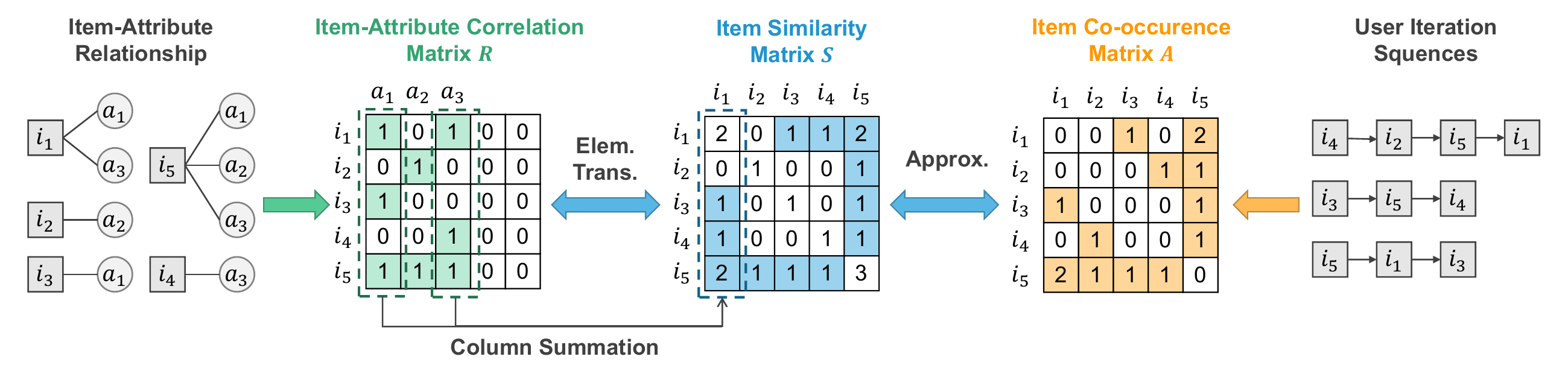}  
  \caption{The relationship between item-attribute correlation matrix and Item co-occurence matrix. For ease of illustration, we consider a specific scenario with only five items and three attributes.}
  \label{matrices}
  \vspace{-0.3cm}
\end{figure*}

\section{PRELIMINARIES}
\subsection{Problem Formulation}
Let $\mathcal{U}$ denotes the set of users and $\mathcal{I}$ denotes the corpus of items. For a given user $u \in \mathcal{U}$, we have a historical interaction sequence $\mathcal{S}^{u}=\left\{i_{1}^{u}, i_{2}^{u}, \ldots, i_{L}^{u}\right\}$ ordered chronologically, where $i_{l}^{u} \in \mathcal{I}$ denotes the $l$-th item the user interacted with and $L$ denotes the maximum length of the user behavior sequence. In the matching stage, the goal of multi-interest recommendation is to retrieve a subset of items from $\mathcal{I}$ that user is likely to interact with, based on multiple interests vector extracted from $\mathcal{S}^{u}$.

\subsection{Multi-interest Framework} 
In this section, we provide an overview of the existing multi-interest recommendation framework.

\subsubsection{Embedding Layer}
Let $\mathbf{E}_{\mathcal{I}} \in \mathbb{R}^{|\mathcal{I}| \times d}$ denotes an item embedding matrix that maps all one-hot item IDs into $d$-dimensional dense embedding vectors. Through embedding look-up, the input user interaction sequence $\mathcal{S}^{u}$, which comprises a series of item IDs, is converted into a matrix $\mathbf{H}$:
\begin{equation}\label{eq1}
\mathbf{H}=\operatorname{Embedding}(\mathcal{S}^{u}; \mathbf{E}_{\mathcal{I}})=\left[\mathbf{e}_{1}^{u}, \mathbf{e}_{2}^{u}, \ldots, \mathbf{e}_{L}^{u}\right] \in \mathbb{R}^{L \times d} 
\end{equation}
where $\mathbf{e}_{l}^{u} \in \mathbb{R}^{1\times d}$ is the embedding of the $l$-th item $i_{l}^{u}$.

\subsubsection{Multi-interest Extraction Module}
The aim of this module is to calculate the item-to-interest weight matrix $\mathbf{W}$ based on the embedded user sequences, and then obtain the multi-interest matrix $\mathbf{V}_{u}$ for the user $u$:
\begin{equation}   
\mathbf{W}=\mathcal{F}(\mathbf{H})\in \mathbb{R}^{K \times L}
\end{equation}
\begin{equation}   
\mathbf{V}_{u}=\mathbf{WH} \in \mathbb{R}^{K \times d} 
\end{equation}
where $\mathcal{F}$ represents the function for obtaining the weight matrix, and $K$ is the number of interest vectors.

\subsubsection{Training and Serving}
In the training phase, given the target item $i^{+}$ and $\mathbf{V}_{u}$, we first utilize the argmax operator to select a interest vector which most relevant to the target item:
\begin{equation}
\mathbf{v}_{u}=\mathbf{V}_{u}\left[\operatorname{argmax}(\mathbf{V}_{u} \mathbf{e}_{i^{+}}^{\top}), :\right] \in \mathbb{R}^{1 \times d}
\end{equation}
where $\mathbf{e}_{i^{+}}$ is embedding of the $i^{+}$. Based on the selected user interest $\mathbf{v}_{u}$, the likelihood of user-item interaction $P(i^+ \mid u)$ can be calculated as follows:
\begin{equation}
P(i^{+} \mid u) = \frac{\operatorname{exp}(\mathbf{v}_{u}\mathbf{e}_{i^{+}}^{\top})}{\operatorname{exp}(\mathbf{v}_{u}\mathbf{e}_{i^{+}}^{\top}) + \sum_{{i^{-}} \in \mathcal{N}}\operatorname{exp}(\mathbf{v}_{u}\mathbf{e}_{i^{-}}^{\top})}
\end{equation}
To alleviate the computational complexity resulting from the large number of items, the sampled softmax \cite{sampled-softmax} is employed here, with $\mathcal{N}$ denoting the set of randomly sampled negative samples shared in a batch. The loss function is formulated as follows to maximize $P(i^+ \mid u)$: 
\begin{equation}
\mathcal{L}(\theta) = \sum_{u \in \mathcal{U}} -\log {P(i^{+} \mid u)}
\end{equation}

For the online serving phase, we adopt the fast nearest neighbors algorithm (\textit{e.g.}, Faiss \cite{faiss}) to retrieve the top-N most relevant items for each interest vector from the large-scale item corpus. Then we calculate the relevance scores $f(u, i)$ between the user and these K×N candidate items:
\begin{equation}
f(u, i)=\max _{1 \leq k \leq K}(\mathbf{v}_{u}^{(k)} \mathbf{e}_{i}^{\top})
\end{equation}
where $\mathbf{v}_{u}^{(k)}$ is the $k$-th interest of user $u$. Finally, we use the $f(u, i)$ to reselect the top-N items as the final recommendation results.

\section{INVESTIGATION OF ATTRIBUTE SIMULATION FROM INTERACTION DATA}\label{theory}
Within the multi-interest recommendation framework, there is a frequent need to calculate item similarity with simple functions like dot product on item embeddings. Therefore, a well-clustered item embedding space plays a crucial role in improving the matching performance of the model. Although attribute has proven to be effective side information for item embedding enhancement, the laborious task of annotating results in their unavailability or incompleteness in many public datasets. To address this challenge, we theoretically explore the potential of simulating item-attribute correlations from user-item interaction in this section.

Assume we have a complete attribute set $\mathcal{A}$. By using one-hot ID encoding from 1 to $|\mathcal{I}|$ for items and similarly from 1 to $|\mathcal{A}|$ for attributes, the item $i_{i} \in \mathcal{I}$ denotes the item with ID $i$, and $a_{j} \in \mathcal{A}$ denotes the attribute with ID $j$. As shown in Figure \ref{matrices}, we can depict the relationship between items and attributes as a binary item-attribute correlation matrix $\mathbf{R}\in \mathbb{R}^{|\mathcal{I}| \times |\mathcal{A}|}$, which represents items in rows and attributes in columns. The element $r_{ij}=1$ signifies item $i_{i}$ possesses attribute $a_{j}$, while $r_{ij}=0$ indicates the opposite.

For the item $i_{i}$, we can obtain the ID set of its attributes as: 
\begin{equation}
\mathcal{C}_i=\left\{j \mid r_{i, j} \neq 0, 1 \leq j \leq |\mathcal{A}|\right\}
\end{equation}
Consider a matrix $\mathbf{S}\in \mathbb{R}^{|\mathcal{I}| \times |\mathcal{I}|}$, where the $i$-th column is equal to the sum of the column vectors of matrix $\mathbf{R}$ with indices in $\mathcal{C}_i$:
\begin{equation}\label{col_plus}
\mathbf{S}\left[:, i\right]=\sum_{j \in \mathcal{C}_i}\mathbf{R}\left[:, j\right] \in \mathbb{R}^{|\mathcal{I}| \times 1 }
\end{equation}
Upon applying Equation (\ref{col_plus}) to all columns of matrix $\mathbf{S}$, it is evident that $\mathbf{S}$ is derived via elementary column transformations and matrix extension on $\mathbf{R}$, which can formulate as:
\begin{equation}
\mathbf{S}=\left[\mathbf{R} : \mathbf{O}\right]\mathbf{P} 
\end{equation}
where [:] denotes the horizontal concatenation of two matrices. $\mathbf{O}\in \mathbb{R}^{|\mathcal{I}| \times (|\mathcal{I}|-|\mathcal{A}|)}$ is a zero matrix, and $\mathbf{P} \in \mathbb{R}^{|\mathcal{I}| \times |\mathcal{I}|} $ is an elementary matrix. 

The off-diagonal elements in $\mathbf{S}$ depict the similarity between items. For example, as shown in Figure \ref{matrices}, the element $s_{15}=2$ indicates a certain degree of similarity between item $i_{1}$ and $i_{5}$ with two shared attributes, while $s_{12}=0$ indicates no similarity between item $i_{1}$ and $i_{2}$ due to the absence of shared attributes. Hence, matrix $\mathbf{S}$ also serves as an item similarity matrix. 

The item similarity naturally prompts us with the co-occurrence of items in the user interaction sequence. As the consecutive appearance of two items suggests their potential similarity and a higher co-occurrence frequency indicates a stronger similarity, the item co-occurrence matrix can also reflect the similarity between items to some extent. Suppose $\mathbf{A}\in \mathbb{R}^{|\mathcal{I}| \times |\mathcal{I}|}$ is an item co-occurrence matrix, where element $\mathrm{a_{ij}}$ represents the number of times item $i_{i}$ and item $i_{j}$ co-occur. After normalizing, $\mathbf{A}$ can be viewd as a statistical approximation of the item similarity matrix $\mathbf{S}$. Furthermore, the following formula can be derived:
\begin{equation}\label{eq11}
\left[\mathbf{R} : \mathbf{O}\right] = \mathbf{S}\mathbf{P}^{-1} \approx \mathbf{A}\mathbf{P}^{-1}
\end{equation}
where $\mathbf{P}^{-1} \in \mathbb{R}^{|\mathcal{I}| \times |\mathcal{I}|}$ is the inverse of elementary matrix $\mathbf{P}$. 
This formula demonstrates that by applying reverse elementary column transformations to matrix $\mathbf{A}$, we can obtain an approximation of the item-attribute correlation matrix $\mathbf{R}$ without manual annotation.

\section{METHOD}
The findings in Section \ref{theory} provide us with an inspiring theoretical feasibility of attribute simulation. However, applying it in the multi-interest learning of the matching stage is not trivial due to the need for both high simulation accuracy and low complexity. In this section, we introduce our solution and propose a simple yet effective item embedding enhancement module.

\subsection{Construction of Item Co-occurrence Matrix}
Before training, we first create an item co-occurrence matrix $\mathbf{A} \in \mathbb{R}^{|\mathcal{I}| \times |\mathcal{I}|}$ initializing as a zero matrix. By iterating through all user sequences in the training set, we execute the following operation on the matrix element $\mathrm{a_{ij}}$ when item $i_{i}$ and item $i_{j}$ co-occur:
\begin{equation}
\mathrm{a_{ij}} = \begin{cases}
                  \mathrm{a_{ij}}, & \text{if } T - d_{ij} < 0 \\
                  \mathrm{a_{ij}} + T - d_{ij}, & \text{otherwise}
                \end{cases}
\end{equation}
where $d_{ij}$ is the step interval between $i_{i}$ and $i_{j}$ and T is the threshold for the step interval. As the co-occurrence matrix is symmetric, the same operation is also applied to $\mathrm{a_{ji}}$. Here, we assign higher values to pairs of closer items, emphasizing the higher likelihood of similarity between them. Moreover, we set the diagonal elements of matrix A to 1 and proceed with row normalization.

\subsection{Item Embedding Enhancement via Simulated Attribute}\label{derivation}
To tackle the problem of inadequate item embeddings, our goal is to integrate attribute side information for enhancement.

Ideally, we would have complete attribute information for all items in the item-attribute correlation matrix $\mathbf{R}$. Let $\mathbf{E}_{\mathcal{A}} \in \mathbb{R}^{|\mathcal{A}| \times d}$ denotes the embedding matrix of attributes and $\mathbf{e}_{a_j} \in \mathbb{R}^{1 \times d}$ denotes the embedding of attribute $a_j$ obtained by querying $\mathbf{E}_{\mathcal{A}}$. We can derive the enhanced item embedding $\mathbf{e}_{i_i}$ of item $i_i$ as:
\begin{equation}\label{eq13}
\mathbf{e}_{i_i} = \sum_{j=1}^{|\mathcal{A}|} r_{ij} \mathbf{e}_{a_j}   \in \mathbb{R}^{1 \times d}
\end{equation}
where $r_{ij}$ is the element in the item-attribute correlation matrix $\mathbf{R}$. It is important to mention that this formula can also add item ID-based embedding. However, in an ideal situation where all attributes are known, just as there are no two identical leaves in the world, there would not be two items with exactly all the same attributes. Therefore, unlike the coarse-grained attributes obtained through manual labeling, when attributes are fine-grained and comprehensive, relying solely on item attributes suffices for item representation, and there will not be duplicate item embeddings. Consequently, adding item ID-based item embeddings is redundant.

We can further implement the Equation (\ref{eq13}) in matrix form as:
\begin{equation}\label{eq14}
\mathbf{E}_{\mathcal{I}} = \mathbf{R}\mathbf{E}_{\mathcal{A}} = \left[\mathbf{R} : \mathbf{O}\right]\mathbf{E}^{\prime}_{\mathcal{A}}
\end{equation}
where $\mathbf{E}_{\mathcal{I}} \in \mathbb{R}^{|\mathcal{I}| \times d}$ denotes the enhanced item embedding matrix, and $\mathbf{E}^{\prime}_{\mathcal{A}} \in \mathbb{R}^{|\mathcal{I}| \times d}$ is obtained by vertically expanding matrix $\mathbf{E}_{\mathcal{A}}$. Since $\mathbf{R}$ is unavailable, following Equation (\ref{eq11}), we can further substitute with the co-occurrence matrix $\mathbf{A}$ as:
\begin{equation}\label{eq15}
\mathbf{E}_{\mathcal{I}} = \mathbf{A}\mathbf{P}^{-1}\mathbf{E}^{\prime}_{\mathcal{A}}
\end{equation}

Since the elementary matrix $\mathbf{P}^{-1}$ is unidentified, we can treat it as a parameter matrix. However, as $\mathbf{P}^{-1}$ is a square matrix with a high dimension of $\mathcal{I}$, it will bring a notable increase in the model's parameter count, and performing multiplications between such sizable matrices also results in unbearable computational complexity. 

Facing these challenges, it is necessary to further simplify the formula. According to the associative law of matrix multiplication, we can rewrite Equation (\ref{eq15}) as follows:
\begin{equation}\label{eq16}
\mathbf{E}_{\mathcal{I}} = \mathbf{A} \left(\mathbf{P}^{-1}\mathbf{E}^{\prime}_{\mathcal{A}}\right)
\end{equation}
We can observe that $\mathbf{P}^{-1}\mathbf{E}^{\prime}_{\mathcal{A}}$ actually conducts row elementary transformations on the embedding matrix $\mathbf{E}^{\prime}_{\mathcal{A}}$. Hence, we can merge the process by directly introducing a unified parameter matrix $\widetilde{\mathbf{E}} \in \mathbb{R}^{|\mathcal{I}| \times d}$ to represent the embedding matrix $\mathbf{E}^{\prime}_{\mathcal{A}}$ after elementary transformations. As a result, Equation (\ref{eq16}) can simplify as:
\begin{equation}\label{eq17}
\mathbf{E}_{\mathcal{I}} =  \mathbf{A} \widetilde{\mathbf{E}}
\end{equation}
Interestingly, we can observe a formal resemblance between Equation (\ref{eq14}) and (\ref{eq17}), which suggests that we can regard the item co-occurrence matrix $\mathbf{A}$ as a surrogate item-attribute co-occurrence matrix $\mathbf{R}$, and $\widetilde{\mathbf{E}}$ takes on the role of a dimension-aligned attribute embedding matrix. Through their multiplication, we can concurrently accomplish attribute simulation and attribute-weighted summation, resulting in an enhanced item embedding matrix.

In summary, we discard the existing item ID-based embedding layer and proposed the SimEmb module, with overall workflow as follows:

1) Perform a multiplication between the item co-occurrence matrix and the attribute embedding matrix to obtain an enhanced item embedding matrix $\mathbf{E}_{\mathcal{I}}$, as described in Equation (\ref{eq17}).

2) Embed the input user sequence $\mathcal{S}^{u}$ by performing embedding-lookup on $\mathbf{E}_{\mathcal{I}}$, as described in Equation (\ref{eq1}). 

\subsection{Complexity Analysis and Process Optimization}
In this section, we analyze the additional time complexity resulting from the replacement of the item ID-based embedding layer with the SimEmb module, and further optimize the module's process. Our analysis leads to the following conclusions: During the training phase, the SimEmb module introduces an acceptable increase in time complexity; during the serving phase, the SimEmb module does not introduce any additional time complexity.

In the training phase, the additional time complexity primarily originates from Equation (\ref{eq17}). Considering that the item co-occurrence matrix $\mathbf{A}$ takes the sparse matrix form and assuming its density is $\rho$ (detailed values are presented  in Table \ref{datasets}), the resultant time complexity is $O(\rho|\mathcal{I}|^{2}d)$. This complexity is impractical due to the huge $|\mathcal{I}|$ in the matching stage. Therefore, we further optimize the process of SimEmb as follows:

1) Perform the embedding-lookup on $\mathbf{A}$ to yield the correlation matrix $\mathbf{C}$ between items in the user sequence $\mathcal{S}^{u}$ and their attributes:
\begin{equation}\label{eq18}   
\mathbf{C}=\operatorname{Embedding}\left( \mathcal{S}^{u}; \mathbf{A}\right) \in \mathbb{R}^{L \times |\mathcal{I}|} 
\end{equation}

2) Derive the enhanced item embeddings via weighted summation on  attribute embeddings:
\begin{equation}\label{eq19}
\mathbf{H} = \mathbf{C} \widetilde{\mathbf{E}}=\left[\mathbf{e}_{1}^{u}, \mathbf{e}_{2}^{u}, \ldots, \mathbf{e}_{L}^{u}\right] \in \mathbb{R}^{L \times d} 
\end{equation}

Following the optimized process, the additional time complexity primarily originates from Equation (\ref{eq19}) and can be notably reduced to $O(\rho L |\mathcal{I}| d)$, which is demonstrated to be totally acceptable by experiments in Section \ref{Time Comparison}.

In the serving phase, it's important to highlight that there is no need to operate attribute-weighted summation as we can preserve the learned enhanced item embedding matrix by operating Equation (\ref{eq17}) once and directly conduct embedding-lookup on it. Therefore, there is no additional time complexity introduced. 

\begin{table}[t]
  \caption{Dataset Statistics.}
  \label{datasets}
  \begin{tabular}{lcccc}
    \toprule
    {Datasets} & {\#Users} & {\#Items} & {\#Interactions}& {Density}\\
    \midrule
    {Beauty} & {22,363} & {12,101} & {198,502} & {0.0734\%}\\  
    {Books} & {603,668} & {367,982} & {8,898,041} & {0.0040\%}\\
    {RetailRocket} & {37,175} & {25,310} & {440,084} & {0.0468\%}\\
    {Yelp} & {19,242} & {14,142} & {201,237} & {0.0740\%}\\
    \bottomrule
  \end{tabular}
\end{table}

\begin{table*}[t]
    \setlength{\abovecaptionskip}{0.1cm}
    \caption{Model comparison. Bold: best performance. Underlined: suboptimal performance.}
    \label{Overall Performance}
  \begin{tabular}{cc|cccccccc|cc}
    \toprule
    Datasets & Metric & PopRec & Y-DNN & GRU4Rec & MIND & ComiRec & PIMIRec & Re4 & REMI &  SimRec & Improv.\\
    \midrule
    \multirow{6}{*}{Beauty} 
    & {R@20} &{0.0236} & {0.0643} & {0.0480} & {0.0822} & {0.0650} & {0.0731} & {0.0741} & \underline{0.0779} & \textbf{0.1139} & {+46.21\%}\\
    & {R@50} &{0.0473} & {0.1025} & {0.0768} & {0.1259} & {0.1102} & {0.1101} & {0.1142} & \underline{0.1259} & \textbf{0.1810} & {+43.76\%}\\
    & {ND@20} &{0.0173} & {0.0526} & {0.0414} & {0.0650} & {0.0415} & {0.0477} & \underline{0.0539} & {0.0537} & \textbf{0.0758} & {+40.63\%}\\
    & {ND@50} &{0.0260} & {0.0642} & {0.0502} & {0.0769} & {0.0554} & {0.0631} & {0.0665} & \underline{0.0687} & \textbf{0.0927} & {+34.93\%}\\
    & {HR@20} &{0.0411} & {0.1135} & {0.0845} & {0.1390} & {0.1100} & {0.1305} & {0.1229} & \underline{0.1346} & \textbf{0.1873} & {+39.15\%}\\
    & {HR@50} &{0.0845} & {0.1748} & {0.1310} & {0.1989} & {0.1775} & {0.2070} & {0.1873} & \underline{0.2088} & \textbf{0.2758} & {+32.09\%}\\    
    \midrule
    \multirow{6}{*}{Books}
    & {R@20} &{0.0144} & {0.0380} & {0.0295} & {0.0636} & {0.0525} & {0.0723} & {0.0602} & \underline{0.0768} & \textbf{0.0937} & {+22.01\%}\\
    & {R@50} &{0.0246} & {0.0620} & {0.0472} & {0.0937} & {0.0819} & {0.1107} & {0.0994} & \underline{0.1123} & \textbf{0.1408} & {+25.38\%}\\
    & {ND@20} &{0.0129} & {0.0323} & {0.0291} & {0.0639} & {0.0366} & {0.0510} & {0.0552} & \underline{0.0636} & \textbf{0.0715} & {+12.42\%}\\
    & {ND@50} &{0.0171} & {0.0419} & {0.0365} & {0.0746} & {0.0484} & {0.0652} & {0.0700} & \underline{0.0765} & \textbf{0.0877} & {+14.64\%}\\
    & {HR@20} &{0.0318} & {0.0834} & {0.0642} & {0.1323} & {0.1088} & {0.1449} & {0.1305} & \underline{0.1532} & \textbf{0.1799} & {+17.43\%}\\
    & {HR@50} &{0.0535} & {0.1323} & {0.1026} & {0.1897} & {0.1670} & {0.2144} & {0.2067} & \underline{0.2189} & \textbf{0.2607} & {+19.10\%}\\
    \midrule
    ~ & {R@20} &{0.0190} & {0.3096} & {0.2762} & {0.3797} & {0.4133} & {0.4179} & \underline{0.4631} & {0.4537} & \textbf{0.4859} & {+7.10\%}\\
    ~ & {R@50} &{0.0319} & {0.3896} & {0.3318} & {0.4753} & {0.4959} & {0.5110} & {0.5385} & \underline{0.5483} & \textbf{0.6012} & {+9.65\%}\\
    {Retail} & {ND@20} &{0.0161} & {0.2597} & {0.2637} & {0.3281} & {0.3709} & {0.3643} & {0.3580} & \textbf{0.3877} & \underline{0.3812} & {-0.17\%}\\
    {Rocket} & {ND@50} &{0.0202} & {0.2760} & {0.2732} & {0.3406} & {0.3829} & {0.3757} & {0.3677} & \textbf{0.3990} & \underline{0.3946} & {-0.11\%}\\
    ~ & {HR@20} &{0.0379} & {0.4519} & {0.4139} & {0.5519} & {0.5906} & {0.5912} & {0.6226} & \underline{0.6283} & \textbf{0.6600} & {+5.05\%}\\
    ~ & {HR@50} &{0.0605} & {0.5468} & {0.4806} & {0.6455} & {0.6772} & {0.6807} & {0.6934} & \underline{0.7160} & \textbf{0.7577} & {+5.82\%}\\
    \midrule
    \multirow{6}{*}{Yelp}
    & {R@20} &{0.0257} & {0.0549} & {0.0244} & {0.0624} & {0.0625} & \underline{0.0842} & {0.0598} & {0.0725} & \textbf{0.0921} & {+27.03\%}\\
    & {R@50} &{0.0474} & {0.1034} & {0.0470} & {0.1373} & {0.1282} & \underline{0.1487} & {0.1183} & {0.1328} & \textbf{0.1881} & {+41.64\%}\\
    & {ND@20} &{0.0194} & {0.0462} & {0.0212} & {0.0479} & {0.0547} & \underline{0.0644} & {0.0489} & {0.0638} & \textbf{0.0679} & {+5.43\%}\\
    & {ND@50} &{0.0280} & {0.0650} & {0.0300} & {0.0764} & {0.0787} & \underline{0.0861} & {0.0695} & {0.0852} & \textbf{0.0984} & {+15.49\%}\\
    & {HR@20} &{0.0468} & {0.1179} & {0.0468} & {0.1319} & {0.1340} & \underline{0.1668} & {0.1210} & {0.1616} & \textbf{0.1839} & {+13.80\%}\\
    & {HR@50} &{0.0914} & {0.2135} & {0.0919} & {0.2655} & {0.2556} & \underline{0.2784} & {0.2270} & {0.2732} & \textbf{0.3345} & {+22.44\%}\\
    \bottomrule
  \end{tabular}
\end{table*}

\section{EXPERIMENTS}
\subsection{Experimental Setting}

\subsubsection{Datasets}
We conduct experiments on four benchmark datasets:

\begin{itemize}
    \item \textbf{Amazon Beauty} and \textbf{Books}\footnote{https://nijianmo.github.io/amazon}\cite{amazon} contain user reviews from Amazon.com. We select two widely used sub-categories.
    \item \textbf{RetailRocket}\footnote{https://www.kaggle.com/datasets/retailrocket/ecommerce-dataset} is a real-world e-commerce dataset widely used for research in personalized recommendation. 
    \item \textbf{Yelp}\footnote{https://www.yelp.com/dataset} is a dataset of user-generated reviews and ratings for businesses collected from the Yelp platform. We only keep the transaction records from January 1st, 2019, to December 31st, 2019.
\end{itemize}
We preprocess the dataset according to \cite{comirec}, treating all interactions as implicit feedback. Users or items with fewer than five occurrences are removed. The maximum sequence length is 20. The preprocessed statistics are presented in Table \ref{datasets}.

\subsubsection{Baselines}
Following \cite{comirec,re4,remi}, we compare SimRec with two types of matching baselines: general models and multi-interest models. CF-based and graph-based matching methods are excluded due to the need to model unseen users in the evaluation protocol.
\begin{itemize}
    \item \textbf{PopRec} is a traditional method that recommends users with the most popular items.
    \item \textbf{YouTube DNN} \cite{y-dnn} adopts a two-tower structure and employs average pooling to extract user representation.
    \item \textbf{GRU4Rec} \cite{gru4rec} is an RNN-based model applying GRU to capture complex temporal patterns in user sequences.
    \item \textbf{MIND} \cite{mind} is a pioneer multi-interest work applying the dynamic routing mechanism with the capsule network.
    \item \textbf{ComiRec-SA} \cite{comirec} employs a multi-head self-attention mechanism for multiple interest extracting. We choose the SA version, which is more robust than the DR version.
    \item \textbf{PIMIRec} \cite{pimirec} is a multi-interest model that utilizes the periodicity and interactivity in the sequence.
    \item \textbf{Re4} \cite{re4} is the first work introducing the backward flow into multi-interest learning.
    \item \textbf{REMI} \cite{remi} is an advanced multi-interest model that improves negative sample sampling and routing regularization.
\end{itemize}

\subsubsection{Evaluation Protocal \& Metrics}
\footnotetext[4]{\label{ndcg}NDCG is computed according to ComiRec official revised code (https://github.com/THUDM/ComiRec), which adheres to the paper's NDCG calculation description but differs from the results in the original paper. }

To ensure fairness, we follow the evaluation protocol of prior works \cite{comirec, remi}. Based on the user count, the user sequences in datasets are partitioned into training, validation, and test sets with an 8:1:1 ratio. The entire user sequences in the training set are used for training, while in validation and testing, the first 80\% of user sequences are used to model user representations, and the remaining 20\% serve as target items. To evaluate the performance of our proposed methods, we employ commonly used metrics, including \textbf{Recall}, \textbf{Hit Rate}, and \textbf{NDCG}\textsuperscript{\ref{ndcg}} (Normalized Discounted Cumulative Gain), calculated based on the top 20/50 matched candidates, following \cite{comirec, re4, remi, pimirec}.  

\subsubsection{Implementation Details}
We implement our work with PyTorch 2.0 in Python 3.10. We build SimEmb on ComiRec-SA by default and name it \textbf{SimRec}, except for section \ref{Compatibility}. The batch size is 128 for Books dataset and 256 for other datasets. The embedding dimension is 64, and the number of interests is 4. We use Adam \cite{adam} for optimization with a learning rate 0.001. The maximum training iterations are set to 1 million rounds. The step interval threshold $T$ for Yelp is 5, while other datasets are set at 3.
We test baseline models using the open-source code provided in the original paper and reproduce REMI with PyTorch 2.0. We followed the original paper's settings for other hyperparameters of baselines. For a fair comparison, we set the number of negative samples for all models to 10{*}batch size.

\begin{table}[t]
    \setlength{\abovecaptionskip}{0.1cm}
    \caption{Compatibility results. Performance of advanced multi-interest models and their SimEmb-enhanced version.}
    \label{compatibility experiments}
    \resizebox{\columnwidth}{!}{
    \begin{tabular}{cc|cc|cc|cc}
    \toprule
    {Datasets} & {Metric} & PIMIRec & +SimEmb & Re4 & +SimEmb & REMI & +SimEmb\\    
    \midrule
    \multirow{6}{*}{Beauty} 
    & {R@20} & {0.0731} & {\textbf{0.1082}} & {0.0741} & {\textbf{0.1010}} & {0.0779} & {\textbf{0.1190}}\\
    & {R@50} & {0.1101} & {\textbf{0.1745}} & {0.1142} & {\textbf{0.1708}} & {0.1259} & {\textbf{0.1859}}\\
    & {ND@20} & {0.0477} & {\textbf{0.0689}} & {0.0539} & {\textbf{0.0704}} & {0.0537} & {\textbf{0.0846}}\\
    & {ND@50} & {0.0631} & {\textbf{0.0867}} & {0.0665} & {\textbf{0.0897}} & {0.0687} & {\textbf{0.1027}}\\
    & {HR@20} & {0.1305} & {\textbf{0.1761}} & {0.1229} & {\textbf{0.1663}} & {0.1346} & {\textbf{0.1949}}\\
    & {HR@50} & {0.2070} & {\textbf{0.2655}} & {0.1873} & {\textbf{0.2646}} & {0.2088} & {\textbf{0.2897}}\\    
    \midrule
    \multirow{6}{*}{Books}
    & {R@20} & {0.0723} & {\textbf{0.1031}} & {0.0602} & {\textbf{0.0728}} & {0.0768} & {\textbf{0.1013}}\\
    & {R@50} & {0.1107} & {\textbf{0.1558}} & {0.0994} & {\textbf{0.1217}} & {0.1123} & {\textbf{0.1547}}\\
    & {ND@20} & {0.0510} & {\textbf{0.0809}} & {0.0552} & {\textbf{0.0608}} & {0.0636} & {\textbf{0.0854}}\\
    & {ND@50} & {0.0652} & {\textbf{0.0986}} & {0.0700} & {\textbf{0.0780}} & {0.0765} & {\textbf{0.1031}}\\
    & {HR@20} & {0.1449} & {\textbf{0.1986}} & {0.1305} & {\textbf{0.1494}} & {0.1532} & {\textbf{0.2022}}\\
    & {HR@50} & {0.2144} & {\textbf{0.2877}} & {0.2067} & {\textbf{0.0.2377}} & {0.2189} & {\textbf{0.2926}}\\
    \midrule
    ~ & {R@20} & {0.4179} & {\textbf{0.4822}} & {0.4631} & {\textbf{0.5306}} & {0.4537} & {\textbf{0.4900}}\\
    ~ & {R@50} & {0.5110} & {\textbf{0.5923}} & {0.5385} & {\textbf{0.6310}} & {0.5483} &  {\textbf{0.5979}}\\
    {Retail} & {ND@20} & {0.3643} & {\textbf{0.3860}} & {0.3580} & {\textbf{0.3947}} & {0.3877} & {\textbf{0.3910}}\\
    {Rocket} & {ND@50} & {0.3757} & {\textbf{0.3989}} & {0.3677} & {\textbf{0.4072}} & {0.3990} & {\textbf{0.4036}}\\
    ~ & {HR@20} & {0.5912} & {\textbf{0.6533}} & {0.6226} & {\textbf{0.6824}} & {0.6283} & {\textbf{0.6635}}\\
    ~ & {HR@50} & {0.6807} & {\textbf{0.7515}} & {0.6934} & {\textbf{0.7714}} & {0.7160} & {\textbf{0.7569}}\\
    \midrule
    \multirow{6}{*}{Yelp}
    & {R@20} & {0.0842} & {\textbf{0.1092}} & {0.0598} & {\textbf{0.1055}} & {0.0725} & {\textbf{0.1114}}\\
    & {R@50} & {0.1487} & {\textbf{0.2005}} & {0.1183} & {\textbf{0.1942}} & {0.1328} & {\textbf{0.2169}}\\
    & {ND@20} & {0.0644} & {\textbf{0.0876}} & {0.0489} & {\textbf{0.0787}} & {0.0638} & {\textbf{0.0871}}\\
    & {ND@50} & {0.0861} & {\textbf{0.1159}} & {0.0695} & {\textbf{0.1069}} & {0.0852} & {\textbf{0.1198}}\\
    & {HR@20} & {0.1668} & {\textbf{0.2130}} & {0.1210} & {\textbf{0.1984}} & {0.1616} & {\textbf{0.2229}}\\
    & {HR@50} & {0.2784} & {\textbf{0.3621}} & {0.2270} & {\textbf{0.3429}} & {0.2732} & {\textbf{0.3896}}\\
    \bottomrule
  \end{tabular}
  }
\end{table}

\subsection{Overall Performance}
In this section, we aim to answer the question: How does SimRec perform compared to state-of-the-art matching models?

Table 2 displays the overall performance comparison on four datasets , and we have the following observations :
\begin{itemize}
     \item General sequential recommendation models (Y-DNN and GRU4Rec) surpass PopRec, indicating that modeling the user interest based on the interaction sequence meets personalized recommendation requirements.
     \item Classic multi-interest models (MIND and ComiRec) outperform Y-DNN and GRU4Rec which only extract a single user representation vector, indicating the efficacy of multi-interest frameworks in representing users' various interests reflected by the item diversity in the user sequence.   
     \item Advanced multi-interest models (PIMIRec, Re4, and REMI) achieve better performance than the classic models, with REMI standing out. Despite the primary focus of REMI not being item embedding enhancement, embeddings of irrelevant items are alienated as it revises the negative sampling, demonstrating the significance of a robust embedding space.
     \item SimRec, which applies our proposed SimEmb on the basic ComiRec,  achieves the best performance across nearly all metrics on four datasets. Particularly, in the pivotal Recall@50 metric for the matching stage, SimRec improves baseline performances of Beauty, Books, RetailRocket, and Yelp by 46.21\%, 25.38\%, 9.65\%, and 41.64\%, respectively. While we do not lead in a few metrics, the difference from the best performance is negligible. The results reinforce the merits of item embedding enhancement with simulated attributes. 
\end{itemize}

\begin{table}[t]
    \setlength{\abovecaptionskip}{0.1cm}
    \caption{Comparison of annotated and simulated attributes.}
    \label{Real Attribute}
    \resizebox{\columnwidth}{!}{
    \begin{tabular}{cc|c|cc|cc}
    \toprule
    {Datasets} & {Metric} & {ComiRec} & {+MaaEmb} & {Improv.} & {+SimEmb} & {Improv.}\\
    \midrule
    \multirow{6}{*}{Beauty} 
    & {R@20} & {0.0650} & {0.0741} & {+14.00\%} & \textbf{0.1139} & {+75.23\%}\\
    & {R@50} & {0.1102} & {0.1309} & {+18.78\%} & \textbf{0.1810} & {+64.25\%}\\
    & {ND@20} & {0.0415} & {0.0448} & {+7.95\%} & \textbf{0.0758} & {+82.65\%}\\
    & {ND@50} & {0.0554} & {0.0612} & {+10.47\%} & \textbf{0.0927} & {+67.33\%}\\
    & {HR@20} & {0.1100} & {0.1265} & {+15.00\%} & \textbf{0.1873} & {+70.23\%}\\
    & {HR@50} & {0.1775} & {0.2056} & {+15.83\%} & \textbf{0.2758} & {+55.38\%}\\ 
    \midrule
    \multirow{6}{*}{Books}
    & {R@20} & {0.0525} & {0.0584} & {+11.24\%} & \textbf{0.0937} & {+78.48\%}\\
    & {R@50} & {0.0819} & {0.0913} & {+11.48\%} & \textbf{0.1408} & {+71.92\%}\\
    & {ND@20} & {0.0366} & {0.0412} & {+12.57\%} & \textbf{0.0715} & {+95.36\%}\\
    & {ND@50} & {0.0484} & {0.0542} & {+11.98\%} & \textbf{0.0877} & {+81.20\%}\\
    & {HR@20} & {0.1088} & {0.1201} & {+10.39\%} & \textbf{0.1799} & {+65.35\%}\\
    & {HR@50} & {0.1670} & {0.1844} & {+10.42\%} & \textbf{0.2607} & {+56.11\%}\\
    \midrule
    ~ & {R@20} & {0.4133} & {0.4308} & {+4.23\%} & \textbf{0.4859} & {+17.57\%}\\
    ~ & {R@50} & {0.4959} & {0.5408} & {+9.08\%} & \textbf{0.6012} & {+21.23\%}\\
    {Retail} & {ND@20} & {0.3709} & {0.3630} & {-2.13\%} & \textbf{0.3812} & {+2.78\%}\\
    {Rocket} & {ND@50} & {0.3829} & {0.3767} & {-1.62\%} & \textbf{0.3946} & {+3.06\%}\\
    ~ & {HR@20} & {0.5906} & {0.6057} & {+2.56\%} & \textbf{0.6600} & {+11.75\%}\\
    ~ & {HR@50} & {0.6772} & {0.7079} & {+4.53\%} & \textbf{0.7577} & {+11.89\%}\\
    \midrule
    \multirow{6}{*}{Yelp}
    & {R@20} & {0.0625} & {0.0678} & {+8.48\%} & \textbf{0.0921} & {+47.36\%}\\
    & {R@50} & {0.1282} & {0.1410} & {+9.98\%} & \textbf{0.1881} & {+46.72\%}\\
    & {ND@20} & {0.0547} & {0.0508} & {-7.13\%} & \textbf{0.0620} & {+13.35\%}\\
    & {ND@50} & {0.0787} & {0.0779} & {-1.02\%} & \textbf{0.0984} & {+25.03\%}\\
    & {HR@20} & {0.1340} & {0.1361} & {+1.57\%} & \textbf{0.1839} & {+37.24\%}\\
    & {HR@50} & {0.2556} & {0.2732} & {+6.89\%} & \textbf{0.3345} & {+30.87\%}\\
    \bottomrule
  \end{tabular}
  }
  \vspace{-0.5cm}
\end{table}

\subsection{Compatibility Study}\label{Compatibility}
Since compatibility is one of the key factors limiting the applicability of a method, in this section, we aim to answer the question: Apart from ComiRec, how does SimEmb perform when implemented on other multi-interest models?

We compare the performance of advanced multi-interest models, PIMIRec \cite{pimirec}, Re4 \cite{re4}, and REMI \cite{remi}, and their SimEmb-enhanced version. To ensure fairness, we maintain the other modules and hyperparameters in the original models unchanged and only replace the item embedding layer with the SimEmb. The comparison between the original and enhanced models is presented in Table \ref{compatibility experiments}. 

\begin{figure*}[t]  
    \setlength{\abovecaptionskip}{0.1cm}
    \centering
    \includegraphics[width=\linewidth]{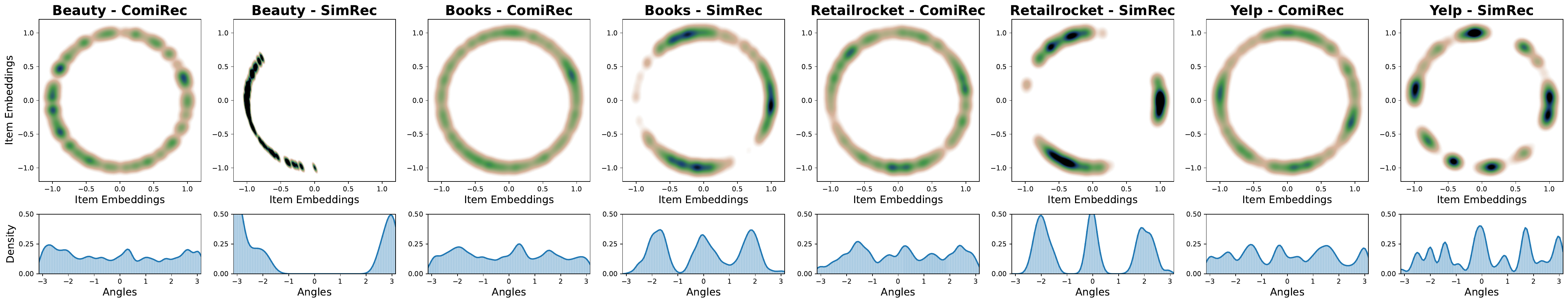}
    \caption{Item embedding distributions of ComiRec and its SimEmb-enhanced version on four datasets. Darker colors indicate denser point clustering in this area and density curve displays point densities across angles.}
    \label{visual fig}
\end{figure*}

It is evident that SimEmb notably improves the performance of all models by an average of 35.85\% on Recall@20 and 38.42\% on Recall@50, demonstrating the strong compatibility of SimEmb. Moreover, the substantial performance improvement across all multi-interest models further indicates their inadequate item embedding space and affirms the effectiveness of our proposed item embedding enhancement solution.

\subsection{Performance Comparison with Manually Annotated Attribute}
In this section, we aim to answer the question: How does SimEmb perform compared to utilizing manually annotated attributes?

We treat the manually annotated categories in the meta dataset as item attributes. By preserving all categories without filtering, we collect 249, 2760, 3718, and 1925 attributes from Beauty, Books, RetailRocket, and Yelp datasets, respectively. To utilize these manually annotated attributes, we embed them into embedding vectors. Then we sum the item-ID embedding with embeddings of the item's annotated attributes to obtain the enhanced item embedding. We refer to this comparison experiment as MaaEmb.

The results in Table \ref{Real Attribute} demonstrate that manually annotated attributes effectively improve the matching performance by enhancing item embeddings. However, SimEmb still surpasses, which is reasonable as the attribute information in the dataset is coarse-grained and incomplete. For example, in the Books dataset, 47.47\% of items (equivalent to 286565 items) are only labeled with 'Books', which essentially equates to the absence of any category information, leading to a limited performance improvement. In contrast, SimEmb bypass manual annotation and directly simulates complete and fine-grained attribute information from interaction data, resulting in more pronounced improvement.

\subsection{Hyperparameter Study}
\begin{figure}[t]  
    \setlength{\abovecaptionskip}{0.1cm}
    \centering
    \includegraphics[width=\linewidth]{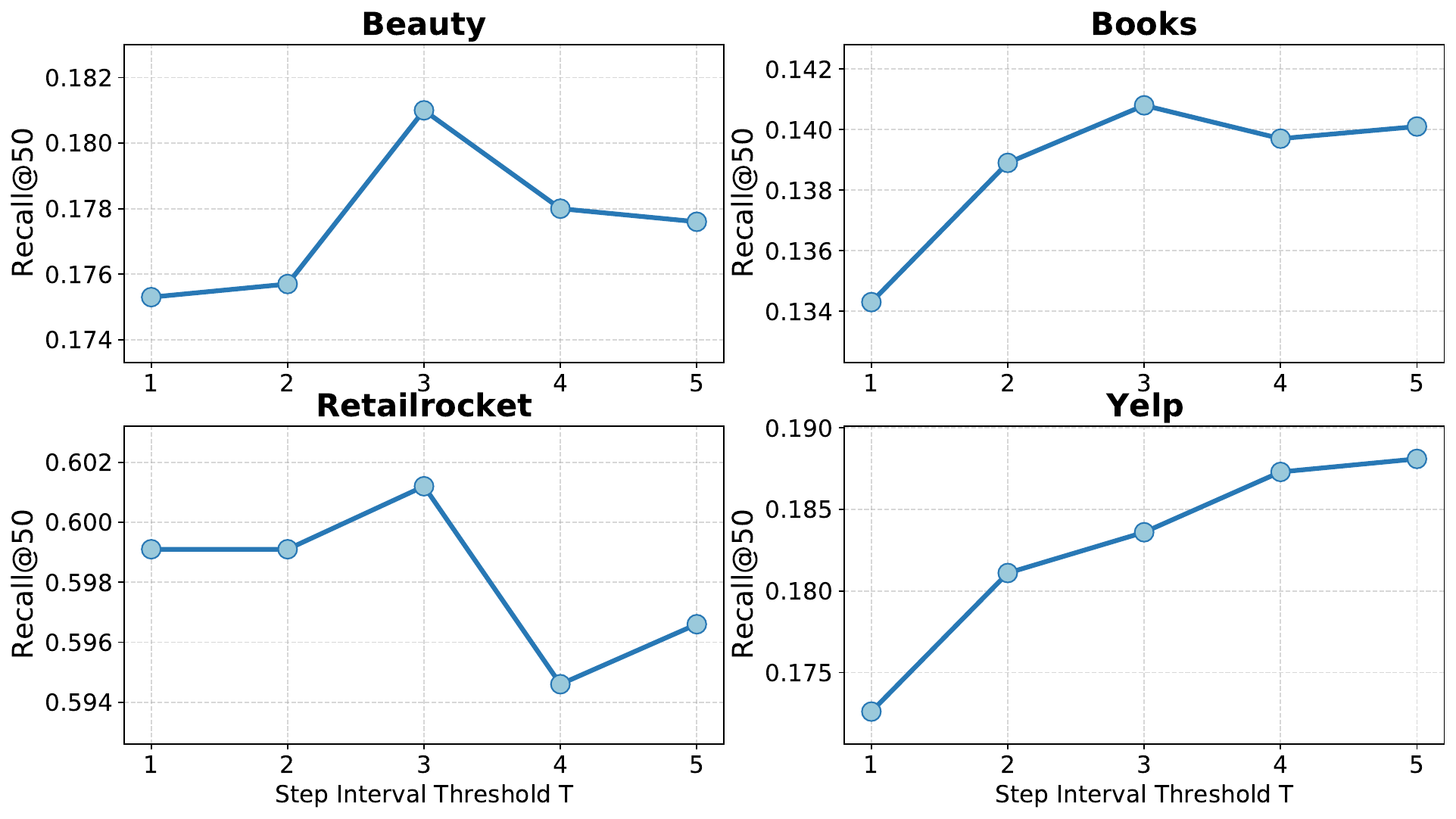}
    \caption{Performance comparison with different T.}
    \label{threshold}
    \vspace{-0.5cm}
\end{figure}

In this section, we investigate the influence of the only hyperparameter in SimEmb, the step interval threshold T. We set T from 1 to 5 to construct different co-occurrence matrices and perform experiment comparisons.

As shown in Figure \ref{threshold}, the Yelp dataset achieves its highest performance at a step interval threshold of 5, while the other datasets perform best with a threshold of 3. Additionally, performance diminishes when the threshold is too small or large, attributed to insufficient statistics at lower thresholds and increased noise at higher thresholds. In both cases, the co-occurrence matrices struggle to capture similarity between items effectively. 
Furthermore, slight fluctuations in model performance are observed when the step interval threshold changes in a proper range, emphasizing the robustness of SimEmb.

\subsection{Training and Testing Time Comparison}\label{Time Comparison}

\begin{table}[t]
    \setlength{\abovecaptionskip}{0.1cm}
    \caption{Training time per batch and testing time (s).}
    \label{Time Comparison tabel}
    \resizebox{\columnwidth}{!}{
    \begin{tabular}{c|cc|cc|cc}
    \toprule
    \multirow{2}{*}{Models} & \multicolumn{2}{c|}{Beauty} & \multicolumn{2}{c|}{Retail} & \multicolumn{2}{c}{Yelp}\\
    ~ & {Train} & {Test} & {Train} & {Test} & {Train} & {Test}\\
    \midrule
    {ComiRec} & {0.0071} & {1.1195} & {0.0091} & {1.4213} & {0.0082} & {1.0509} \\
    {PIMIRec} & {0.0567} & {2.6135} & {0.0726} & {3.3171} & {0.0664} & {2.4637}\\   
    {Re4} & {0.0149} & {1.3676}  & {0.0209} & {1.7926}  & {0.0139} & {1.3463} \\   
    {REMI} & {0.0106} & {1.4773} & {0.0153} & {2.0594} & {0.0127} & {1.3232} \\   
    {SimRec} & {0.0108} & {1.1878} & {0.0179} & {1.4771} & {0.0134} & {1.0851}\\  
    \bottomrule
    \end{tabular}
    }
    \vspace{-0.5cm}
\end{table}

In this section, we report the training time on a batch and the test time on the entire test set of different multi-interest matching models. The results are collected on Intel(R) Xeon(R) Gold 5115 CPU and Quadro RTX 6000 GPU. The training time is obtained by averaging the total training time of the first 500 batches. For all models, we set the batch size as 256.

The results presented in Table \ref{Time Comparison tabel} indicate that ComiRec has the shortest training and testing times due to its lightweight architecture. While other models that incorporate improvements within the ComiRec framework all exhibit corresponding increases in training and testing times. Among them, PIMIRec takes the longest time consumption due to its addition of a complex graph structure. On the contrary, SimRec introduces a lower complexity during the training phase, resulting in an acceptable increase in training time given its superior performance. Moreover, since we can preserve the enhanced item embedding matrix of SimeEmb after training, there is no additional time complexity introduced during testing. As a result, the testing time of SimRec is nearly on par with ComiRec.

\subsection{Visualization Case Study}\label{visualization}

In this section, we present visualizations of the item embedding space in ComiRec and SimRec across four datasets, aiming to showcase the effectiveness of SimEmb intuitively.

For each dataset, we randomly select 600 items from three categories. Following the visualization method in \cite{visual}, we first map the learned item embedding vectors to 2-dimensional normalized vectors on the unit hypersphere with t-SNE \cite{t-sne}. Then, we employ nonparametric Gaussian kernel density estimation (KDE) \cite{kde} to plot the embedding distributions in $\mathbb{R}^2$ and von Mises-Fisher (vMF) KDE to plot the density curve of embedding distributions on angles.

As shown in Figure \ref{visual fig}, the item embeddings of ComiRec exhibit uniform distributions across the circle, with relatively flat density curves. It suggests the clustering deficiency within the item embedding space, which limits the model's matching performance. In contrast, the item embeddings of SimRec form clusters in narrow arcs accompanied with sharper density curves, showcasing a notable improvement of clustering in the item embedding space. This improvement can be attributed to the weighted summation of simulated attributes, where items with shared attributes tend to have similar embeddings, automatically promoting clustering.

\section{RELATED WORK}
\subsection{Deep Candidate Matching}
With the rapid advancement of deep learning, lightweight neural networks have also gained widespread utilization in the matching stage. 
Initial methods grounded in collaborative filtering (CF) \cite{DBLP:conf/kdd/KabburNK13, DBLP:conf/www/SarwarKKR01, DBLP:journals/dsonline/LindenSY03}. Neural CF \cite{ncf} harnessed matrix factorization components and a non-linear multi-layer perceptron (MLP) to effectively model the matching between users and items. As the field progressed, graph-based recommendation \cite{DBLP:conf/sigir/LiCL0DZLT21, DBLP:conf/kdd/PerozziAS14, berg2017graph, DBLP:conf/www/TangQWZYM15, DBLP:conf/kdd/GroverL16, Kipf2016SemiSupervisedCW, DBLP:conf/nips/HamiltonYL17} and sequential recommendation \cite{DBLP:conf/kdd/HuangSSXZPPOY20, DBLP:conf/recsys/YiYHCHKZWC19, DBLP:conf/www/YangYCHLWXC20} are introduced into the matching stage. The graph-based model amasses insights from neighbor nodes within the user-item graph to refine the node representation of users and items. LightGCN \cite{lightgcn} and SGC \cite{SGC} streamline the GCN structure and redundant operations, showcasing an efficient and effective approach that inspired subsequent graph-based recommendation models like SGL \cite{SGL}. The sequential recommendation explore the dynamic user interest information inherent within the historical interaction sequence. YouTube-DNN \cite{y-dnn} is an early pioneer in sequential recommendation, employing a dual-tower architecture and deriving user characteristics via average pooling of interacted items. Drawing from natural language processing (NLP), subsequent works adopt frameworks such as recurrent neural networks \cite{gru4rec} and memory networks \cite{DBLP:conf/wsdm/ChenXZT0QZ18}.

\subsection{Multi-interest Recommendation}
Recent studies in sequential recommendation have found that extracting a unified user representation vector from sequences is insufficient to capture the multiple interests of the user. As this concern is more prominent in the matching stage, multi-interest recommendations have gained increasing attention. MIND \cite{mind} pioneers using a dynamic routing mechanism to extract multiple interest vectors from the user sequence. ComiRec \cite{comirec} introduces a framework based on the multi-head attention mechanism and harmonizes the trade-off between recommendation diversity and precision. PIMIRec \cite{pimirec} considers the periodicity and interactivity within sequences. UMI \cite{umi} and UDM \cite{udm}, respectively, leverage user profiles and target items to assist in extracting multiple interest vectors. Re4 \cite{re4} introduces the backward flow to reevaluate interest embeddings. REMI \cite{remi} questions the uniformly sampled softmax and routing collapse in the multi-interest framework, proposing an improved negative sampling strategy and the routing regularization method. Despite these works making improvements in the multi-interest extraction module and loss functions from various perspectives, few efforts are devoted to enhancing item embeddings, which limits the effectiveness of item matching.

\subsection{Attribute Information Completion in Recommendation}
In practical scenarios, the attribute side information of items may be incomplete or unavailable, posing challenges for attribute fusion in the recommendation. Traditional methods for side information completion fill the missing attribute with heuristic values like average or randomized values \cite{biessmann2018deep, lee2018impute, shi2019adaptive}. However, these context-independent values can perturb the training of model parameters. Another approach is to estimate the values of the missing attribute. Early models based on auto-encoders (AE) \cite{beaulieu2017missing, pereira2020reviewing, DBLP:conf/emnlp/WangNL18} apply random dropout to the side information and reconstruct the lost data. KTUP \cite{cao2019unifying} predicts missing relationships within the knowledge graph to introduce knowledge into the recommendation. CC-CC \cite{shi2019adaptive} calculates the importance of features and proposes an adaptive feature sampling strategy to enhance model robustness when dealing with missing features, avoiding direct prediction of missing data. AGCN \cite{wu2020joint} cyclically updates the graph using approximated attribute values to boost attribute inference and item recommendation. MIIR \cite{lin2023modeling} unifies the tasks of missing attribute information completion and sequential recommendation.

In contrast to our work, these methods have two limitations: 1) The lack of specific design for the matching stage leading to the issue of high complexity; 2) The attribute information completion depends on manually annotated attributes and cannot handle cases where attribute information is unavailable.

\section{CONCLUSION}
In this work, we pinpoint the inadequate item embedding of prior multi-interest models in the matching stage. Building upon the theoretical feasibility, we propose a simple yet effective module, SimEmb, which employs co-occurrence matrices to simulate attribute information for item embedding enhancement. Extensive experiments and visualization demonstrate the superiority of SimEmb in both accuracy and complexity. 
In the future, we will tackle the limitations of our work by focusing on denoising methods during the construction of the co-occurrence matrix, as the unavoidable introduction of noise interferes with the approximation of item similarity. 

We believe our work offers a simple and reliable alternative for the labor-intensive task of manually annotating attributes with broader implications beyond multi-interest recommendation, and can extend to other recommendation fields, thereby inspiring future research.

\bibliographystyle{ACM-Reference-Format}
\bibliography{ref}

\section*{ETHICAL CONSIDERATIONS}

We acknowledge the importance of ethical considerations and potential societal impacts, and our method doesn't exhibit apparent negative consequences. We assure stringent data anonymization and compliance with privacy regulations. Our method's design emphasizes responsible deployment, striving to minimize biases and protect user privacy in alignment with ethical principles.

\end{document}